\documentstyle[aps,twocolumn]{revtex}     
\newcommand{\be}[1]{\begin{equation}\label{#1}}
\newcommand{\ee}{\end{equation}}     
\newcommand{\bea}{\begin{eqnarray}}
\newcommand{\eea}{\end{eqnarray}} 
\newcommand{\eq}[1]{(\ref{#1})}
\newcommand{\EE}{\cal E}
\newcommand{\SS}{\cal S}
\newcommand{\pdif}[2]{\frac{\partial #1}{\partial #2}}
\begin{document}  
%
\title{ \large\bf Time Dependence in Quantum Mechanics}
\author{John S.  Briggs  \footnote{Permanent address: -- Theoretische Quantendynamik -- 
 Fakult\"at f\"ur Physik, Universit\"at Freiburg,
 Hermann--Herder--Str.  3, D--79104 Freiburg,  Germany} 
and Jan M. Rost  \footnote{Permanent address: 
 Max-Planck-Institute for Physics of Complex Systems, N\"othnitzer 
 Str.~38, D-01238 Dresden, Germany
 }
}
\address{Institute for Advanced Study, Wallotstr.~19, D-14193 Berlin, 
Germany\\ \rm (February 1999)}
\author{
\begin{minipage}{152mm} 
\vspace*{5mm} 
It is shown that the time-dependent equations (Schr\"odinger and 
Dirac) for a quantum system can be always derived from the 
time-independent equation for the larger object of the system 
interacting with its environment, in the limit  that the dynamical 
variables of the environment can be treated semiclassically. The time 
which describes the quantum evolution is then provided parametrically 
by the classical evolution of the environment variables. The method 
used is a generalization of that known for a long time in the field 
of ion-atom collisions, where it appears as a transition from the 
full quantum mechanical
{\it perturbed stationary states} to the {impact parameter} method in
which the projectile ion beam is treated classically.
\draft
\end{minipage}
}
\maketitle
\section{Introduction}
Originally Schr\"odinger \cite{Sch26} proposed his wave equation in time independent 
form as an eigenvalue equation for the time-independent Hamilton operator, i.e.
\be{tise}
(E-H)\Psi = 0,
\ee
the time-independent Schr\"odinger equation (TISE).  Subsequently, 
this was generalized to consider quantum systems as being described by 
state vectors in Hilbert space and 
observables by Hermitian operators.  
Pairs of canonically conjugate operators satisfy Heisenberg 
commutation rules and thereby fulfill corresponding Heisenberg 
uncertainty relations.  In a later paper \cite{Sch26a} Schr\"odinger 
introduced a time-dependent Schr\"odinger equation (TDSE) and this 
version has found much wider application than the TISE even for 
time-independent Hamiltonians.  On the one hand this is due to the 
technical reason that it is easier to solve the initial value problem of 
the TDSE than the boundary value problem of the TISE.  On the other 
hand there is the deeper reason that our experience based on the 
classical world  still conditions us to think in terms of 
physical processes proceeding from some initial state and developing 
in time to some final state.  This is despite the fact that quantum 
mechanics teaches that the state $\Psi$ of a closed system can be 
written as a time-independent superposition 
\be{conf}
\Psi = \sum_{n}c_{n}\Phi_{n}
\ee
of states of the same total energy but different internal 
configurations where the probability of measurement of each possible 
state is given by $|c_{n}|^{2}$. 

Almost all books on quantum mechanics simply {\it postulate} a TDSE 
for a quantum system $\SS$ with Hamiltonian $H_{\SS}$ 
\be{tdse}
\left(i\hbar\frac{\partial}{\partial t} - 
H_{\SS}\right)\tilde\psi_{\SS}(x,t) = 0,
\ee
where $\{x\}$ are the system variables. 
Notwithstanding the intuitive appeal of the TDSE there are several 
problems connected with its use, for example;
\begin{itemize}
    \item[(A)] Although many authors have contrived to define one 
    \cite{AhBo61}, there is no simple obvious definition of a Hilbert space 
    operator corresponding to time. The "usual replacement"
    \be{repl}
    E\to i\hbar\frac{\partial}{\partial t} 
    \ee
    is unusual in that $E$ is the eigenvalue of the operator $H$, but 
    $i\hbar\partial/\partial t$ is not the eigenvalue of an operator.
    \item[(B)] Following from (A), there is no energy-time 
    uncertainty relation, although time, as measured by 
     clocks, is arguably an observable and  countless books refer to the 
    ``energy-time'' uncertainty relation.
    \item[(C)] The TDSE admits time-dependent Hamiltonians. Time 
    dependent potentials are often introduced in an ad hoc way and 
    lead to non-Hermitian Hamiltonians. 
    This implies a loss of 
    norm of the quantum wavefunction. i.e.\ to an ad hoc loss of 
    particles from  the quantum system. 
    \item[(D)] The TDSE is simply postulated, with the tacit 
    assumption that the parameter called time is to be identified with 
    the classical time. There is no proof of this. For example, in 
    the case that $H_{\SS}$ is time-independent, corresponding to an 
    isolated conservative system, the simple phase transformation 
    \be{twf}
        \tilde\psi_{\SS}(x,t) = \exp(i/\hbar E_{\SS}t)\psi_{\SS}(x)
    \ee
     in \eq{tdse} leads to the TISE
    \be{tise2}
     (E_{\SS}-H_{\SS})\psi_{\SS}(x) = 0,
    \ee
    where $E_{\SS}$ is well-defined as an eigenvalue of $H_{\SS}$.  
    Clearly, in this procedure (or its inverse), the "time" $t$ 
    merely appears as a mathematical 
    parameter.  
\end{itemize}

Here we will show that these difficulties can all be overcome in that 
the TDSE for a quantum system $\SS$ can be derived from the TISE for 
the larger object comprising the system $\SS$ and its quantum 
environment $\EE$.  In the limit that the environment is so large 
(precisely what this means depends upon the object under discussion) 
that $\cal E$ is practically unchanged in its interaction with $\SS$, 
it can be treated {\it semiclassically}.  In this limit the variables 
of the environment undergo a time development described by classical 
equations involving a classical time parameter as measured by a clock.  
In the same approximation the system $\SS$ develops in time governed 
by the TDSE with an effective time-dependent Hamiltonian whose time 
dependence arises from the interaction with $\EE$ via the implicit 
time-dependence of the classical environment variables.  In this way 
time always arises in quantum mechanics as an externally defined 
classical parameter and time-dependent Hamiltonians from the 
interaction with a classical environment.  The most important example 
of the interaction of a quantum system $\SS$ with an environment is 
the act of observation or measurement, when time is defined by the 
classical (macroscopic) measuring device.  Furthermore, since all 
measurements ultimately involve the detection of charged particles, 
photons, or heat (phonons) these are the types of environment we shall 
consider.  Their classical motion is described by Newton or Maxwell 
equations and hence the time parameter introduced into the TDSE for 
the quantum system is {\it identical} with that entering the classical 
equations.

In deriving a TDSE from the TISE of a composite system, we will follow 
closely the development of Briggs and Macek \cite{BrMa91}, who 
considered the particular case of a beam of ions (environment) 
interacting with a target atom (quantum system).  In this particular 
case they showed how the initially time-independent equation for the 
coupled ion-atom reduces to a time-dependent equation for the atom 
alone, in the limit that the ion is considered to move along a 
classical trajectory.  In fact the method is much older in origin in 
atomic collision physics and can be traced through the "perturbed 
stationary states" (PSS) method of Mott and Massey \cite{Mot65} to the 
original 1931 paper of Mott \cite{Mot31}, where he showed the 
essential equivalence of the time-independent PSS and time-dependent 
impact parameter approaches in the limit of high beam velocities, 
with the time defined by the variables of the {\it 
particle beam}.  The PSS method is a generalization of the adiabatic 
Born-Oppenheimer method as applied to stationary states of molecules.  
Interestingly, this method of molecular physics has become very 
popular recently in defining time in quantum gravity.  Here time is 
introduced into the time-independent Wheeler-de Witt equation by 
treating gravitation semi-classically but the matter field quantum 
mechanically, in a procedure that is similar to that used in the case 
of atomic collisions \cite{BrVe89}.

Starting with the TISE of \eq{tise} for $\EE \otimes \SS$ we will {\it 
derive} the TDSE \eq{tdse} for $\SS$.  Thereby we will show that the 
{\it parametric} time derivative arises from the expectation values of 
the environment {\it operators}, thus resolving problem (A) and 
eliminating the need for the replacement \eq{repl}.  Similarly, we can 
show that the energy-time "uncertainty relation" for $\SS$ arises from 
true (operator based) uncertainty relation for $\EE$, 
explaining problem $(B$).  Furthermore, it will be shown that the 
time-dependence of $H_{\SS}(t)$ arises in a well defined way from the 
interaction of $\SS$ with $\EE$ (problem (C)) and that the time which 
arises is precisely the time describing the classical motion of $\EE$ 
i.e.\ the classical environment provides the clock for the quantum 
system (problem (D)).  The further approximation that the interaction 
of $\SS$ with $\EE$ can be neglected gives a time-independent 
Hamiltonian $H_{\SS}$ appropriate to a non-interacting (closed) 
quantum system and the TDSE \eq{tdse} reduces to the new TISE 
\eq{tise2} for $\SS$ alone.

It is clear that the procedure can now be continued in that $\SS$, 
described by the TISE \eq{tise2}, can  be considered as composed of 
$\SS'$ and $\EE'$ to define time and a TDSE for $\SS'$.  The extent to 
which the subdivision is valid depends on the accuracy with which the 
dynamics of $\EE'$ may be approximated (semi-)classically.  More 
pragmatically, one might say that the position of the interface 
between the quantum and the classical worlds depends upon the degree 
of precision set (or achieved) by the measurement.

In section II the interaction of a quantum system with a material 
environment is considered.  The TDSE is derived by generalizing the 
procedure due to Briggs and Macek \cite{BrMa91} who considered the 
particular case of an atom interacting with a particle beam.  Then it 
is shown how the energy-time uncertainty relation arises.  In section 
III the general procedure is illustrated by the three generic examples 
of a system $\SS$ interacting with a particle beam or with a set of 
quantum oscillators (photons or phonons) as environment $\EE$.  
Finally, the same procedure can be applied to transform the 
time-independent Dirac equation (TIDE) into the time-dependent Dirac 
equation (TDDE).  Here it is interesting to observe that the time 
component of the spacetime of the quantum system actually arises from 
the implicit time variation of the {\it classical environment} 
variables, i.e. again it is this variation that provides the clock for 
the quantum system.
\section{The emergence of time}
We begin by decomposing the total Hamiltonian $H$  for the large 
object in \eq{tise} into 
\be{henv}
H = H_{\EE}+H_{\EE\SS}+ H_{\SS}
\ee
with the Hamiltonians $H_{\EE}$ for the environment and $H_{\SS}$ for 
the quantum system.  
For convenience we assume a coordinate representation for $\EE$ with a 
 standard form of $H_{\EE} = K + V_{\EE}$ where the potential energy is a 
 function of coordinates only, $V _{\EE}= V_{\EE}(R)$ and the kinetic energy is 
 written in mass scaled coordinates $R = (R_1,R_{2},\ldots)$ as a sum 
 over all degrees of freedom, $i = 1,\ldots,n$:
 \be{kine}
 K = -\frac{\hbar^{2}}{2M}\sum_{i}^{n}\pdif{^{2}}{R^{2}_{i}}.
\ee
Almost all relevant environments can be cast into this form as will be 
illustrated in section III. 
In \eq{henv} $H_{\EE\SS}$  describes the coupling between $\SS$ 
and $\EE$.  However, as a consequence of the environment 
being ``large'' 
compared to the quantum system, the coupling is  asymmetric in the sense that 
the state $\chi$ of the environment $\EE$ depends negligibly on the 
variables $\{x\}$ of the system $\SS$ while the system state $\psi$ 
depends on the environment variables $\{R\}$.  Accordingly, we write 
the total wavefunction $\Psi$ in \eq{tise} as
 \be{prodw} 
 \Psi(x,R) = \chi(R)\psi(x,R).
 \ee
 Having defined the wavefunction of the system we can express  what
 a ``large'' environment  means in  terms of an 
 {\it asymmetry condition}. It defines and 
 distinguishes, in the decomposition of $H$, environment $\EE$ and 
 system $\SS$ through the respective energy expectation values by
\be{asym}
\langle\chi|H_{\EE}|\chi\rangle_{R} \equiv E_{\EE}\gg E_{\SS}\equiv 
\langle\chi|U_{\SS}|\chi\rangle_{R},
\ee 
where
\be{hsexpec}
U_{\SS}(R) = \langle\psi(x,R)|H-V_{\EE}|\psi(x,R)\rangle_{x}.
\ee
A more detailed discussion of the requirements for the validity of 
\eq{prodw} and a derivation of the form of $U_{\SS}(R)$  in 
\eq{hsexpec} is given in the appendix,  starting from a formally 
exact ``entangled'' wavefunction  $\Psi(x,R) = 
\sum_{n}\chi_{n}(R)\psi_{n}(x,R)$ for the complete object composed of 
system and environment. 
It will  be shown there that
the asymmetry condition \eq{asym}  justifies a posteriori the form 
\eq{prodw} of the wavefunction.  As is well known from adiabatic 
approximations in other contexts a wavefunction of the form \eq{prodw} 
can only be justified a posteriori if a condition such as  \eq{asym}  
is fulfilled. 

 Backed by the asymmetry condition \eq{asym} and \eq{hsexpec} we 
 determine $\chi$ from the eigenvalue equation
 \be{tisee}
 (H_{\EE} +U_{\SS}(R) - E)\chi(R) = 0.
\ee
The term $U_{\SS}$ represents the very small influence $\SS$ has on 
the state of the environment.  The environment is taken to be a large, 
quasiclassical system so that its state vector $\chi$ can be 
approximated by a semiclassical wavefunction
\be{wkb}
\chi(R) = A(R)\exp(iW/\hbar),
\ee
where $W(R,E)$ is the (time-independent) action of the {\it 
classical}  Hamiltonian $H_{\EE}$. 

  Inserting \eq{henv} and \eq{prodw} into 
\eq{tise} we get using
\eq{tisee} 
\bea
\label{scoupl}
&&\chi\left[U_{\SS}(R)-H_{\SS}-H_{\EE\SS}\right]\psi(x,R)\nonumber\\
&&= 
\frac{\hbar}{i}\sum_{i}C_{i}\pdif{\psi(x,R)}{R_{i}},
\eea
where the operator $C_{i}$ is given by
\be{mixed}
C_{i}=\chi\frac{1}{2M}\frac{\hbar}{i}\pdif{}{R_{i}}+
\frac{1}{M}\frac{\hbar}{i}\pdif{\chi}{R_{i}}.
\ee
In accordance with the asymmetry condition \eq{asym} we have assumed
that 
\be{commo}
{}[H_{\EE\SS},\chi] = 0.
\ee
Note that \eq{scoupl} is an equation for the wavefunction $\psi$ while $\chi$, 
already fixed in  \eq{tisee} acts like a potential, i.e. an operator.

With the form of \eq{wkb} for $\chi$ we can write for the operator $C_{i}$ of 
\eq{mixed}
\be{deriv1}
C_{i} = 
\chi\left[\frac{1}{2M}\frac{\hbar}{i}\left(\pdif{}{R_{i}}+\frac{2}{A}
\frac{\partial A}{\partial R_{i}}\right)+ \frac{1}{M}
\pdif{W}{R_{i}}\right]
\ee
where $(\partial W/\partial R_{i})M^{-1} = P_{i}/M = dR_{i}/dt$ is the 
{\it classical} momentum vector of the environment $\EE$.  The most 
important step to turn \eq{scoupl} into a TDSE for the system $\SS$ 
and its wavefunction $\psi$ is to keep only the term of lowest order in 
$\hbar$ in \eq{deriv1} which reduces the operator $C_{i}$ to
\be{Cclas}
C_{i}\approx \chi \frac{1}{M}
\pdif{W}{R_{i}} = \chi \frac{P_{i}}{M} = \chi\frac{dR_{i}}{dt}.
\ee
From this approximation for $C_{i}$ emerges the classical time on the 
right hand side of \eq{scoupl} through
\be{timep}
\sum_{i} C_{i}\pdif{}{R_{i}} = \chi
\sum_{i}\frac{dR_{i}}{dt}\pdif{}{R_{i}}  = \chi\frac{d}{dt}.
\ee
Since the $R_{i}$ are reduced to 
 classical variables $R_{i}(t)$ \eq{scoupl} can now be written
\be{suncoup}
(H_{\SS}+ H_{\EE\SS} -i\hbar 
\left.\pdif{}{t}\right|_{x}-U_{\SS}(t))\psi(x,t) = 0
\ee
Here we emphasize that the time derivative is to be taken with $\{x\}$ 
fixed since it arises from the derivative w.r.t. the independent  
(quantum) variables $\{R\}$. Finally, a phase or gauge transformation
\be{gauge}
\psi = \exp[i/\hbar\int_{-\infty}^{t}(U_{\SS}(t'))dt']\tilde\psi
\ee
leads to the TDSE for the quantum system alone.  To write it in a 
familiar form we might specify, although not necessary, $H_{\EE\SS} = 
V(x,t)$, i.e.  the interaction with the environment is expressed as a 
potential.  Then we have from \eq{suncoup} and \eq{gauge}
\be{tdse2}
(H_{\SS} + V(x,t) - i\hbar\pdif{}{t})\tilde\psi(x,t) = 0.
\ee

Having accepted \eq{tdse2}, the whole structure of time-dependent 
quantum mechanics, e.g., the transition to Heisenberg and interaction 
pictures, time-dependent perturbation theory etc., can be developed as 
usual.  Indeed, Briggs and Macek show explicitly that, in the same 
approximations that lead to \eq{tdse2}, the time-independent T-matrix 
element for the system plus environment reduces to a time-dependent 
transition amplitude for the system alone.  Similarly, a precise 
consideration of the nature of the interaction $V(x,t)$ should allow 
one to derive the many variations of stochastic Schr\"odinger 
equations that have been proposed to model the interaction of a 
quantum system with an environment or measuring device. 

One further observation must be made.  This is the question of the 
``uncertainty relation'' for energy and time.  Since there is no 
canonical operator for time there is no uncertainty relation in the 
sense of Heisenberg.  That quoted in many books arises from the basic 
property of the Fourier transform from energy space to time space in 
which time appears as a mathematical, rather than a mechanical 
(physical) variable.  However, within the approximation of the 
environment as a classical object a time-energy relation for the 
quantum system can be {\it derived} from the uncertainty relation for 
the environment, since it is the position variable of the environment 
that defines the classical time.  For any two operators related to the 
environment, one has
\be{abs-unc}
\Delta A\Delta B \ge \frac{1}{2i}\langle[A,B]\rangle_{R}.
\ee
In particular if $A = H_{\EE}$ and $B = R_{i}$, then
\be{HRunc}
\Delta H_{\EE} \Delta R_{i} \ge \hbar/2\frac{\langle P_{i}\rangle}{M}.
\ee
Now, in the classical limit for the environment variables, 
$\Delta R_{i} = v_{i}\Delta t$ and $v_{i} = P_{i}/M = \langle 
P_{i}\rangle/{M}$, so that we obtain
\be{entime}
\Delta E_{\EE}\Delta t \ge \hbar/2.
\ee
However, from \eq{tisee} and \eq{hsexpec} $E = E_{\EE}+E_{\SS}$ where $E$ is the fixed 
total energy. Hence, $\Delta E_{\EE}= \Delta E_{\SS}$ and \eq{entime} 
becomes 
\be{uncertaint}
\Delta E_{\SS}\Delta t \ge \hbar/2,
\ee
i.e. the energy-time  uncertainty for the quantum system emerges from 
the fluctuations in the expectation values of the environment 
variables. 

Note that for the derivation and application of this uncertainty 
relation it is necessary that the quantum system interacts with 
the environment through the potential $H_{\EE\SS}$. It is this 
interaction that leads to the uncertainty in the system energy 
$E_{\SS}$. In the same way the uncertainty in the time $\Delta t$ 
arises from the time development described by \eq{tdse2} where the 
time is the classical time defined by the classical time-development 
of the environment.  Such energy-time relationships are to be 
distinguished from those usually postulated for the {\it isolated} quantum 
system alone, i.e. where the operators are those of the quantum system.
In that case $H_{\EE\SS}= 0$ and the quantum systems satisfy the TISE
\eq{tise2}. A time energy `uncertainty relation'' then only arises by 
introduction of a mathematical time through Fourier transform of 
\eq{tise2} to a ``time'' space. The uncertainties $\Delta E$ and 
$\Delta t$ then refer to the widths of Fourier distributions and the 
uncertainty relation to the inverse relation between them.

\section{Specific examples}
To illustrate the general approach of section II in more detail we 
will consider three examples which represent the  most common 
ways in which quantum systems are probed and measured.
First we will discuss the interaction of the quantum system with a particle 
beam, then we will describe the interaction with a  "bath" of 
oscillators, e.g.  photons or phonons. Finally, we will show that
also in a relativistic environment the time-dependent Dirac equation
(TDDE) can be derived  from the time-independent Dirac equation
(TIDE) in a way analogous to the non-relativistic case.  

\subsection{A particle beam as environment}
A particle beam of fixed momentum $\vec{P}=\hbar\vec{K}$ interacting 
with a quantum system was considered by Briggs and Macek 
\cite{BrMa91}.  The asymmetry condition \eq{asym}, necessary to 
separate environment and quantum system, is achieved when the beam 
kinetic energy $P^{2}/2M$ is much greater than than the energy 
differences $\Delta E_{S}$ in the system states populated as a result 
of the interaction.  The (semi-)classical limit for the environment, 
necessary to justify \eq{wkb}, is reached when $\vec{P}$ is so large 
that the de Broglie wavelength is far shorter than the extent of the 
quantum system.  In this case the WKB-wavefunction for a free 
beam-particle is the exact quantum solution.  We may choose a 
coordinate system with the $z-$axis along the beam direction,
\be{chibeam}
\chi(X,Y,Z) = (2\pi)^{-3/2}\exp(iP_{Z}Z/\hbar),
\ee
where $\vec{P} = (0,0,P_{Z})$ is the classical momentum of the beam.
 Since \eq{chibeam} is of the form \eq{wkb} it
leads  directly to the emergence of time according to 
\eq{timep},
\be{timepbeam}
\frac{P_{Z}}{M}\pdif{}{Z} = \frac{dZ}{dt}\pdif{}{Z} = \frac{d}{dt}.
\ee

\subsection{A quantized field as environment}
In the second example the environment comprises a collection of 
quantum oscillators with mode frequencies $\omega_{k}$, i.e. photons 
or phonons.  Then the Hamiltonian $H_{\EE}$ can be written as a sum  
over field modes
\be{hfield}
H_{\EE} = 
\sum_{k}\left[\frac{P^{2}_{k}}{2m_{k}}
+\frac{\omega_{k}^{2}Q_{k}^{2}}{2}\right],
\ee
where $m_{k}=1$ for photons. The field operators satisfy the usual 
commutation relations
\be{cofield}
[Q_{k},P_{k'}] = i\hbar\delta_{kk'}
\ee
with
\be{poperat}
P_{k} = -i\hbar\pdif{}{Q_{k}}.
\ee
The  energy $E_{\EE}$ is the eigenenergy of \eq{hfield} and $\chi(Q)$
is an eigenvector in the $Q$-representation which need not be 
specified further here. Then, using \eq{prodw}, we write
\be{prodw-field}
\Psi(x,Q) = \chi(Q)\psi(x,Q).
\ee
Substitution of $\Psi$ into \eq{tise} leads as before to \eq{scoupl}.
The quasiclassical field limit  leads now for the operator $C_{i}$ in
\eq{scoupl} to the replacement
\be{Cfield}
C_{i} = \frac{P_{k}}{m_{k}}\chi \to \chi \frac{dQ_{k}}{dt}
\ee
where $Q_{k}(t)$ is now a {\it classical} field amplitude. 
Since \eq{Cfield} is of the same form as \eq{Cclas}  time emerges
as in \eq{timep} and the  TDSE is established for the case of a 
quasi-classical field as environment. 

\subsection{An example of relativistic dynamics}
Finally, we consider how the relativistic generalization of the 
transition from the TISE to TDSE occurs for femions, i.e.  how a 
time-independent Dirac equation (TIDE) for system plus environment 
becomes a time-dependent Dirac equation for the system.  To keep the 
derivation simple we restrict ourselves to a quantum object of two 
fermions whose spins are uncoupled and where the energies are such 
that pair production can be neglected.  In this case the classical 
limit will be where the relativistic mass $M$ of the environment 
fermion becomes much greater than that of the system fermion $m$.  The 
TIDE can be written formally as in \eq{tise} with the Hamiltonian 
\eq{henv} whose elements are now defined as,
\begin{mathletters}
\bea
\label{diracE}
H_{\EE}&=& c\vec \alpha_{\EE}\vec{P}_{\EE} + V_{\EE}(\vec R) + 
\beta_{\EE}Mc^{2}\\
H_{\SS}&=& c\vec \alpha_{\SS}\vec{P}_{\SS} + V_{\SS}(\vec x) + 
\beta_{\SS}mc^{2}.
\label{diracS}
\eea
\end{mathletters}
The potentials $V_{\EE}, V_{\SS}$ are potentials acting separately 
on  environment and system particles, respectively and can be 
neglected in what follows, i.e. we consider two free fermions 
interacting through a coupling Hamiltonian $H_{\EE\SS}$. The total 
wavefunction $\Psi$ is then written as in \eq{prodw} but is now a 
product of a spinor $\chi(\vec{R})$ representing the spin state of 
$\EE$ and a spinor $\psi(\vec{x},\vec{R})$ representing the spin 
state of the system but depending parametrically on the space 
variables of the environment. 

The analogue of \eq{scoupl} now becomes
\be{relcoup}
[\chi(E_{\EE}-E+H_{\SS}+H_{\EE\SS}) - 
c\vec{\alpha}_{\EE}\chi\vec{P}_{\EE}]\psi(\vec{x},\vec{R}) = 0.
\ee
However, the operator $c\vec\alpha_{\EE}$ is just the velocity operator 
\cite{BjDr64}
which for positive energy solutions has the form 
$c^{2}\vec{P}_{\EE}/E_{\EE}$. For free motion the exact solution is 
the same as the semiclassical one.  However Jensen and Bernstein \cite{JeBe84}  
have 
shown that even when potentials as in \eq{diracE}  and \eq{diracS} are 
retained, in lowest order semiclassical approximation the {\it form} 
of the velocity operator is unchanged. Then, in this
 limit, with 
$E_{\EE} = Mc^{2}$ one has
\be{Crel}
\vec{P}_{\EE}/M = \vec v_{\EE} = \frac{d\vec{R}}{dt}
\ee
so that \eq{relcoup} becomes 
\be{relred}
[H_{\SS}+H_{\EE\SS}- 
i\hbar\vec{v}_{\EE}\nabla_{R}]\psi(\vec{x},\vec{R}(t)) =
U_{\SS}\psi(\vec{x},\vec{R}(t))
\ee
or
\be{tdde}
\left[H_{S}+H_{\EE\SS}-i\hbar\left.\pdif{}{t}\right|_{x}\right]\psi(\vec{x},t)
= U_{\SS}(t)\psi(\vec{x},t).
\ee
With the phase transformation \eq{gauge} one has
\be{tdde2}
\left[H_{S}+H_{\EE\SS}-i\hbar\pdif{}{t}\right]\tilde\psi(\vec{x},t)=0,
\ee
the time dependent Dirac equation. Here, it is interesting to note 
that the time coordinate of the {\it quantum system} spacetime arises 
from the space coordinate of the {\it classical environment}.  

\section{Summary}
We began our considerations with a time independent stationary state 
of a complete object comprising system and environment.  The 
semiclassical treatment of the environment $\EE$ with the requirement 
that its own state and energy to zeroth order are unaffected by the 
quantum system $\SS$, has led to a TDSE for this system in which the 
quantum variables of the environment are replaced by classical 
variables.  The interaction with the environment then appears as 
explicitly time-dependent and the motion of the environment provides a 
time derivative which monitors the development of the quantum system.  
If the interaction with the environment is ignored i.e.  the quantum 
system is closed, then time is reduced to a mere mathematical variable 
and can be removed entirely by the simple phase transformation 
\eq{twf} leading to the TISE of \eq{tise2}.  Note, however, that it is 
inconsistent to put $V(x,t) = H_{\EE\SS}$ in \eq{tdse2} to zero in 
order to obtain a TDSE of the form of \eq{tdse} which is simply 
postulated.  Were $H_{\EE\SS}$ zero, the TDSE \eq{tdse2} cannot be 
derived, since the Hamiltonian \eq{henv} is then fully separable in 
$x$ and $R$ and instead of the {\it approximation} \eq{prodw} one has 
an {\it exact} solution of the form $\Psi(x,R) = \chi(R)\psi(x)$.  
This has the consequence that system and environment are fully 
decoupled  implying that the environment can no longer 
provide time  for the system.  Formally, one can see this 
from the uncertainty relation \eq{uncertaint}. As $H_{\EE\SS}\to 0$, 
the energy exchange between environment and system vanishes  
and $\EE$ and $\SS$ separately become isolated without uncertainty in
in their respective energy. Hence, in \eq{uncertaint} with $\Delta 
E_{\SS}\to 0$  we get $\Delta t\to\infty$.  
In 
this sense time arises and is meaningful for a quantum system only 
when interaction with a quasi-classical external environment defines a 
clock (e.g.  an oscillator) with which the time development is 
monitored.
\section*{Acknowledgment}
We would like to thank L.~Diosi for helpful discussions.

\begin{appendix}
    \section{}
In the following we will show under what approximations and with
which consequences an exact or ``entangled'' wavefunction,
\be{sumwf}
\Psi(x,R) = \sum_{n}\chi_{n}(R)\psi_{n}(x,R)
\ee
for the large object described by the Hamiltonian $H$ in \eq{tise}, 
leads to the product wavefunction \eq{prodw}. This form of the 
wavefunction  describes a 
state of $H =
H_{\EE}+H_{\SS}+H_{\EE\SS}$ if    the system $\SS$ and the environment $\EE$ 
are weakly coupled under the asymmetry condition \eq{asym}.
Without loss of generality   we may assume the $\psi_{n}$ to be 
orthonormal for each $R$, i.e. 
$\langle\psi_{n}|\psi_{m}\rangle = \delta_{nm}$, where   here 
and in the following brackets
$\langle|\rangle$ denote integration over system variables $x$ only.

We first proceed exactly as in the Born-Oppenheimer, or better, 
perturbed stationary states (PSS) approximation of molecular physics 
\cite{Mot65} where \eq{sumwf} is substituted in 
$H_{\EE}+H_{\SS}+H_{\EE\SS}-E|\Psi\rangle = 0$ and a projection is 
made onto a particular state $\psi_{m}$ to give
\be{projsum}
\sum_{n}\langle\psi_{m}|H_{\EE}+H_{\SS}+H_{\EE\SS}-E|\psi_{n}\rangle\chi_{n} = 0
\ee
Making use of the explicit form of $H_{\EE}$ given in \eq{kine} we see 
that \eq{projsum} describes a state $\chi_{m}$ of the environment 
'closely coupled'  to all other states $\chi_{n}$:
\bea\label{ccbo}
&& H_{\EE}\chi_{m} + \sum_{n}
\langle\psi_{m}|H_{\SS}+H_{\EE\SS}|\psi_{n}\rangle\chi_{n}\nonumber\\
&&-\sum_{n,i}\left[\langle\psi_{m}|\frac{\hbar^{2}}{2M}
\pdif{^{2}}{R^{2}_{i}}|\psi_{n}\rangle
+\langle\psi_{m}|\pdif{}{R_{i}}|\psi_{n}\rangle\frac{\hbar^{2}}{2M}
\pdif{}{R_{i}}\right]\chi_{n} \nonumber\\
&&= E\chi_{m}
\eea
This is the full quantum equation for the environment whose states 
$\chi_{n}$ are mixed by the "back-coupling" from the system. The first 
set of coupling terms on the lhs. of \eq{ccbo} are called potential 
couplings and usually the $\psi_{n}$ are chosen to diagonalize these 
terms. The remaining terms are the "dynamical couplings", since their 
off-diagonal matrix elements describe changes in the state of the 
quantum system induced by the motion of the environment. Clearly, in 
order to fulfill the conditions that we demand for separation of 
environment and system, it is necessary that all off-diagonal 
couplings are small, i.e. the environment is insensitive to changes 
in the state of the system. Then \eq{ccbo} reduces to the single-channel equation
\be{schan}
(H_{\EE}+E_{m}(R)-E)\chi_{m}(R) = 
\sum_{i}\langle\psi_{m}|\pdif{}{R_{i}}|\psi_{m}\rangle\frac{\hbar^{2}}{2M}
\pdif{\chi_{m}}{R_{i}}
\ee
where
\be{diag}
E_{m}(R) = \langle\psi_{m}|H_{\SS}+H_{\EE\SS}
-\sum_{i}\frac{\hbar^{2}}{2M}
\pdif{^{2}}{R^{2}_{i}}|\psi_{m}\rangle.
\ee
The dynamical coupling involving the second derivatives 
$\pdif{^{2}}{R_{i}^{2}}$ has been incorporated formally in $E_{m}(R)$.  
However, in connection with the semiclassical approximation for 
$\chi_{m}$ which must be made to derive the TDSE for the system, it is 
consistent to neglect this term.  This is shown explicitly in section 
II in the reduction of the operator $C_{i}$.  Note that this forces a 
choice of the environment such that the major $R$-dependence is 
contained in the $\chi_{m}$ and the $\psi_{m}$ are slowly varying 
functions of $R$.  Then, since
\be{diag1}
\langle\psi_{m}|\pdif{^{2}}{R_{i}^{2}}|\psi_{m}\rangle = 
\sum_{n} |\langle\psi_{m}|\pdif{}{R_{i}}|\psi_{n}\rangle|^{2} ,
\ee
the requirement that the term on the lhs. of this equation is small 
ensures that the off-diagonal dynamical couplings in \eq{ccbo} are 
also small.

If the $\psi_{m}$ can be chosen real, the dynamical coupling terms on 
the rhs of \eq{schan} vanish.  If the $\psi_{m}$ are complex, these 
terms give rise only to geometric (or Berry) phases that can  be 
accounted for by a phase transformation of the $\chi_{n}$. 
Effectively, then \eq{schan} reduces to the eigenvalue equation 
\be{vibeq}
(H_{\EE}+E_{m}(R) -E)\chi_{m}(R) = 0
\ee
for the state of the environment when the quantum system is in the 
state  $\psi_{m}$. Note that the environment is still coupled to the system 
in that the different states of the system provide separate 
potential surfaces $E_{m}(R)$ for the motion of the environment. The 
complete independence of the environment from the precise state of the 
system is achieved in the approximation that the {\it differences} in 
the $E_{m}(R)$ can be replaced by an average potential leading to a 
common $R$ dependence for all $\chi_{m}(R)$, i.e., $\chi_{m} = 
a_{m}\chi$, where the $a_{m}$ are constants. This gives the simplified 
form of \eq{sumwf} 
\be{sumwf2} 
\Psi(x,R) = \chi(R)\sum_{n}a_{n}\psi_{n}(x,R) \equiv \chi(R)\psi(x,R)
\ee
corresponding to the ansatz of \eq{prodw}. Similarly \eq{vibeq}  
becomes identical to  \eq{tisee},  where the averaged potential $U_{\SS}$  of 
\eq{hsexpec} assumes the form
\be{hsexpec2}
U_{\SS}(R) = \langle\psi|H-V_{\EE}(R)|\psi\rangle = 
\sum_{m}|a_{m}|^{2}E_{m}(R)
\ee
with $E_{m}(R)$ from \eq{diag}.

Note that the product ansatz \eq{sumwf2} (\eq{prodw} of the text)  and 
the environment equation \eq{tisee} evaluated in the lowest order WKB 
approximation lead directly to the TDSE \eq{tdse2} for the quantum 
system.  The analysis of this appendix shows how the environment must 
be chosen "large enough" so that it is insensitive to the 
back-coupling from the system.  This insensitivity is necessary to 
derive an effective TDSE from the TISE for the composite object of 
system coupled to environment.

\end{appendix}

\end{document}